# Carrier Dynamics in Submonolayer InGaAs/GaAs Quantum Dots

Zhangcheng Xu[a)], Yating Zhang

The Key Lab of Advanced Technique and Fabrication for Weak-Light Nonlinear Photonics Materials(Ministry of Education), Nankai University, Tianjin 300457, P. R. China and National Laboratory for Infrared Physics, Chinese Academy of Sciences, Yutian Road No. 500, Shanghai 200083, P. R. China

Jørn M. Hvam

Dept. of Communications, Optics and Materials, and Nano.DTU, Technical University of Denmark, DK-2800, Lyngby, Denmark

Jingjun Xu

The Key Lab of Advanced Technique and Fabrication for Weak-Light Nonlinear Photonics Materials(Ministry of Education), Nankai University, Tianjin 300457, P. R. China

Xiaoshuang Chen, Wei Lu

National Laboratory for Infrared Physics, Chinese Academy of Sciences, Yutian Road No. 500, Shanghai 200083, P. R. China

Carrier dynamics of submonolayer (SML) InGaAs/GaAs quantum dots (QDs) were studied by micro-photoluminecence (MPL), selectively excited photoluminescence (SEPL), and time-resolved photoluminescence (TRPL). MPL and SEPL show the coexistence of localized and delocalized states, and different local phonon modes. TRPL reveal shorter recombination lifetimes and longer capture times for the QDs with higher emission energy. This suggests that the smallest SML QDs are formed by perfectly vertically correlated 2D InAs islands, having the highest In content and the lowest emission energy, while a slight deviation from the perfectly vertical correlation produces larger QDs with lower In content and higher emission energy.

[a)] Electronic mail: zcxu@nankai.edu.cn

Self-assembled quantum dots (QDs) can be grown either in the conventional Stranski-Krastanow (SK) mode or via submonolayer (SML) deposition [1-5]. SML InGaAs/GaAs QD heterostrutures is verified to be a quantum-dot-quantum-well structure (QDQW), in which local clusters with higher In content are embedded in a lateral quantum well with lower In content.[6,7] Although SML InGaAs QD lasers with high gain or power have been realized recently, [5,8,9] very few works have been carried out on the carrier dynamics of SML QDs, as compared with SK QDs. [1, 10-12] The study of the carrier dynamics of SML QD structures is of great interest not only for the understanding of the fundamental physics of zero-dimensional structures but also for optoelectronic device applications.

In this letter, we use micro-photoluminescence (MPL), selectively excited photoluminescence (SEPL), and time resolved photoluminescence (TRPL) to explore the localization, capture and recombination of carriers in SML QD structures, at low temperature. We found that the capture time and the recombination lifetime of SML QDs depend strongly on the emission energy, which could be explained by analyzing the growth mechanism of SML QDs.

SML InGaAs/GaAs QDs was formed by alternate deposition of 0.5 ML InAs and 2.5 ML GaAs for ten times (see Ref. [6] for the details of the sample preparation). The MPL and SEPL were measured at 10 K, and the TRPL measurements were carried out at 5 K. In MPL, the incident light from a He-Ne laser at the wavelength of 632.8 nm was focused on the sample to a spot of around 2 μm in diameter. In SEPL, a wavelength-tunable Ti:sapphire laser was used for excitation. In the TRPL setup, the sample was cooled in a liquid helium cryostat and excited in the GaAs barriers with 120 fs pulses from a Ti:sapphire laser at the wavelength of 800 nm and the PL signal was collected, dispersed, and synchronously detected using a streak camera with 2.5ps time resolution. The excitation spots in both REPL and TRPL are about 50 μm in diameter.

The size distribution of QD ensembles could vary slightly with the position on the wafer as seen in Fig. 1(a) at low excitation density. The fine structures for the three spectra are different from each other showing emission from individual QDs. At high excitation power density, a peak at 1.326 eV dominates the whole spectrum, and the peak energies for the three excitation points are identical, as shown in Fig. 1(b). This indicates that the peak at 1.326 eV originates from the delocalized states in the studied structure, corresponding to the QW states.[7].

In SEPL measurement with the excitation energy $E_{ex}$ tuned near to the edge of the density of states (DOS) in the QW ($E_{ex}$ = 1.336 eV), a few sharp resonant lines and a resonant PL band appear within the broad PL band, near to one longitudinal optical (LO, 31 ~ 36 meV) and 2LO (66 meV) phonon energies below the excitation energy, respectively, as shown in Fig. 2. To confirm that these sharp lines are not attributed to resonant Raman scattering, the polarization directions of the incident laser beam and the detected PL signal were set to be along the $[110]$ and the $[1,\bar{1},0]$ directions, respectively, in the back-scattering geometry, as in Ref. [13]. A Raman signal cannot be detected in this geometry, according to the selection rules.[14]. When the excitation energy is less than one GaAs LO-phonon energy above the lateral QW ground state in the SML-grown QDQW structure, the probability for the photon-excited carriers (excitons) to relax within the QW states by emission of only longitudinal acoustic (LA) phonons is less than the carrier (exciton) capture probability from QW to QDs by emission of LO phonons. Dots which can be accessed by emission of LO phonons are populated more efficiently, since their delta-function-like DOS can be accessed directly from the excited energy level in the QW by LO phonon emission.

The complex structure in the 1LO resonant peaks consists of several optical phonon modes whose energies are 36.7 meV, 34.5 meV, 32.9meV and 31.3 meV, respectively, as shown in the inset of Fig. 2. We assign these lines to the LO phonon modes in the GaAs barrier, the GaAs/InAs interface, the InGaAs lateral QW, and the InGaAs QDs, respectively. The 2LO resonance occurs at the energy of 66 meV below the excitation energy, nearly two times the LO phonon energies of QW. The co-existence of several optical phonon modes indicates the complex structure of SML QDs.

The electron-hole (e-h) pairs (or excitons) generated in the GaAs barrier are eithercaptured directly into the QW where they relax and are finally captured by the QDs, Or they are directly captured by the QDs. Then the captured carries will recombine inside the QDs. Fig. 3 shows the PL transient of QD states in SML InGaAs/GaAs QD structures, at an excitation density of 101 $W/cm^2$ ( corresponding to $10^{17}$ electron–hole pairs per $cm^3$ per pulse). On the long time-scale (Fig.3a), the PL decay can be well fitted by a mono-exponential function, and the decay time $\tau_d$ can be evaluated. On the short time-scale (Fig.3b), the PL transients can be fitted by the expression [15]

$$I(t) \propto \left[\exp(-t/\tau_r) - \exp(-t/\tau_d)\right]/(\tau_r - \tau_d), \qquad (1)$$

where $\tau_r$ is the rise time of PL transients, which can provide information on carrier capture into the QDs.

Fig. 4 shows the values of $\tau_d$ and $\tau_r$ plotted against the QD emission energy. With increasing QD transition energy, $\tau_d$ decreases from 840 ps to 500 ps, while $\tau_r$ increases from 35 ps to 60 ps. For SK QDs, QDs with higher emission energy are believed to be smaller, and stronger electron-hole overlap occurs inside the QDs, resulting in longer lifetime [16]. Recently, a reduction of the radiative lifetime for smaller SK QDs with higher emission energy has been observed, which was explained by the reduced electron-hole overlap integral due to the larger piezoelectric effect in larger QDs [17] .

However, in the case of SML QDs, the smallest QDs are formed by perfectly vertically correlated 2D InAs islands,[2,3,6,8] and have the highest In content, while slight deviation from the perfect vertical correlation produces larger QDs with lower In content, as schematically shown in Fig. 5. *SML QDs with higher emission energy are larger.* For larger SML QDs, the penetration of the electron-hole wave-functions into the barrier is reduced, leading to larger electron-hole overlap integral and shorter radiative recombination lifetime.

Since thermalization and relaxation processes with the three dimensional GaAs and the two dimensional QW occur on a much faster time scale [18], the measured rise time mainly reflects the capture process into the QD. When the density of carriers generated by each pulse is much higher than the QD density as in the present case, the carrier capture is mediated by Coulomb scattering (Auger process). According to Ref. [19], the Auger coefficient (capture time) decreases (increases) with the increase of the QD diameter, which coincides with the present result.

In summary, we have explored the carrier dynamics of an InGaAs/GaAs QDQW structure formed by submonolayer deposition. The coexistence of the localized states of QDs and the delocalized states of QWs are revealed clearly in the MPL spectra. Different local phonon energies of the QDQW structure are obtained when the excitation energy is tuned close to the edge of DOS of QWs, indicating the complex structure of SML QDs. The recombination lifetime of SML QDs decreases with the increase of QD emission energy. This can be explained by assuming that SML QDs with higher emission energy have lower average In content and larger volume. The Auger carrier capture time for SML QDs increases with the increase of QD volume, which coincides with theoretical predictions.[19]

Enlightening discussions with Vadim Lyssenko and Dan Birkedal are gratefully acknowledged. This work has been supported in part by the National Natural Science Foundation of China (No. 60444010, 60506013), the Danish Technical Science Research Council, the SRF for ROCS (SEM), the Startup fund for new employees of Nankai University, the Natural Science Foundation of Tianjin City, and the PCSIRT.

# Figure captions

Figure 1. Micro-photoluminecence spectra at 10 K at three different points on the wafer (A,B and C) 1mm apart, at low excitation power density (a), and at high excitation power density (b).

Figure 2. The PL spectra of the SML QD structure at the excitation energies of 1.959 eV above the GaAs barrier band gap, and 1.336 eV just above the edge energy of the QW, at 10 K. The inset zooms in the 1LO parts, and the solid line is a guide for the eye.

Figure 3. TRPL detected at different ground states of SML QDs at 5 K, in the long (short) time-scale for the evaluation of the decay (rise) time.

Figure 4 Dependence of the decay (rise) time with respect to the QD emission energy for SML QDs. (a) the integrated PL spectra; (b) the decay time; (c) the rise time.

Figure 5 The schematic diagram showing the relationship between the size and average In content inside SML QDs. The QD regions are circled by the dotted lines. SQD and LQD are referred to as small QDs and larger QDs.

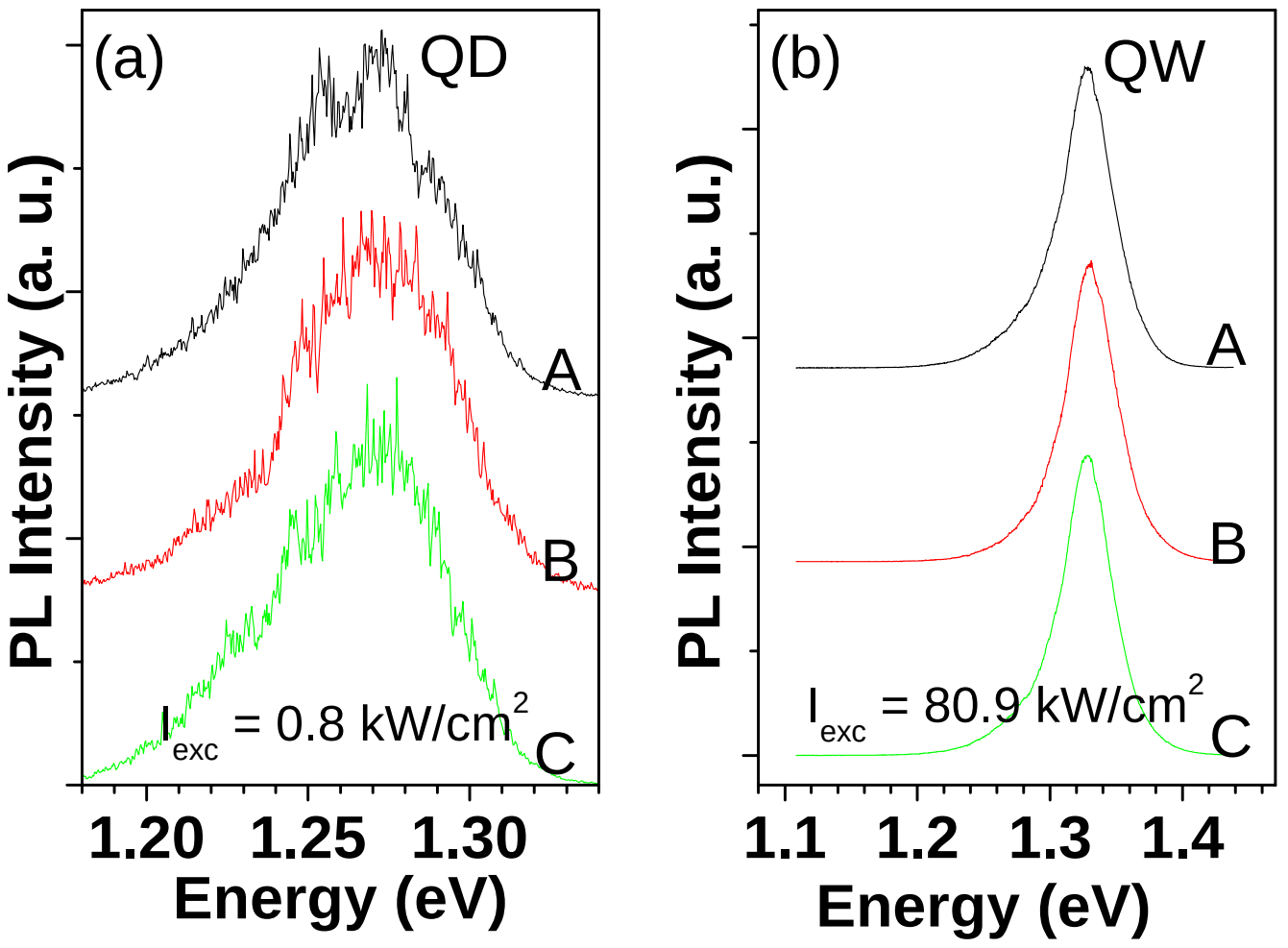

Fig. 1    Zhangcheng Xu et al.

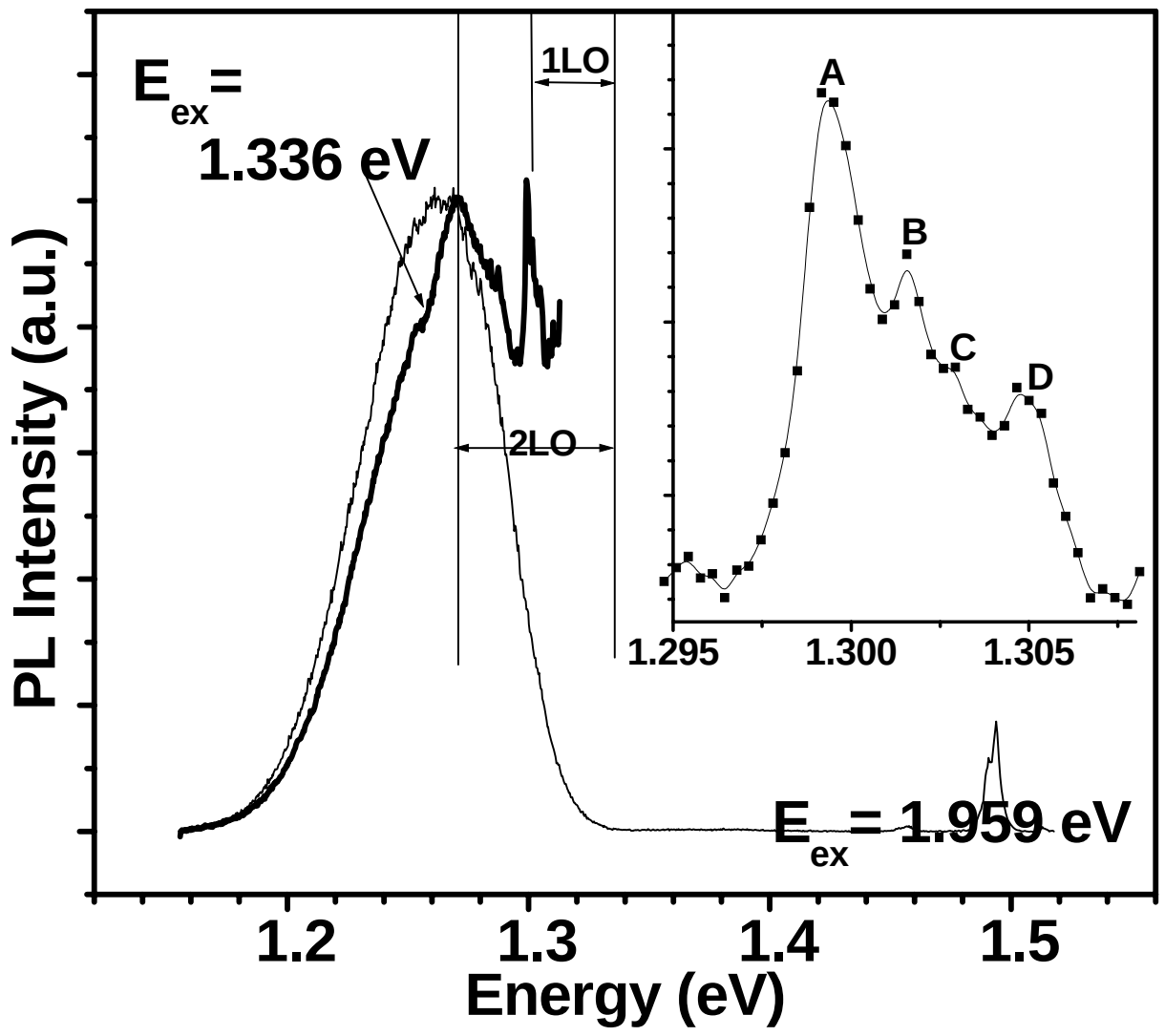

Fig. 2    Zhangcheng Xu et al.

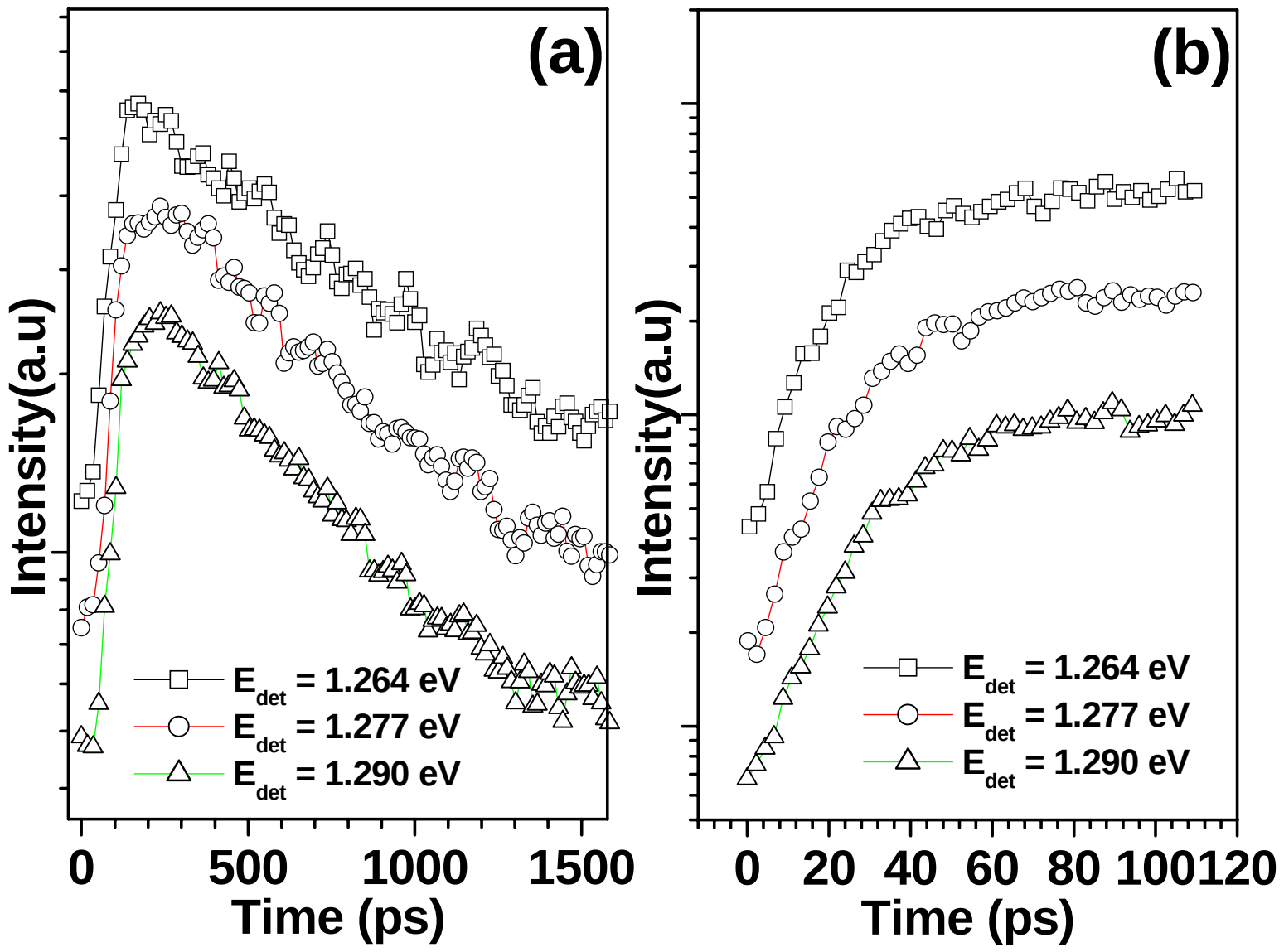

Fig. 3 Zhangcheng Xu et al.

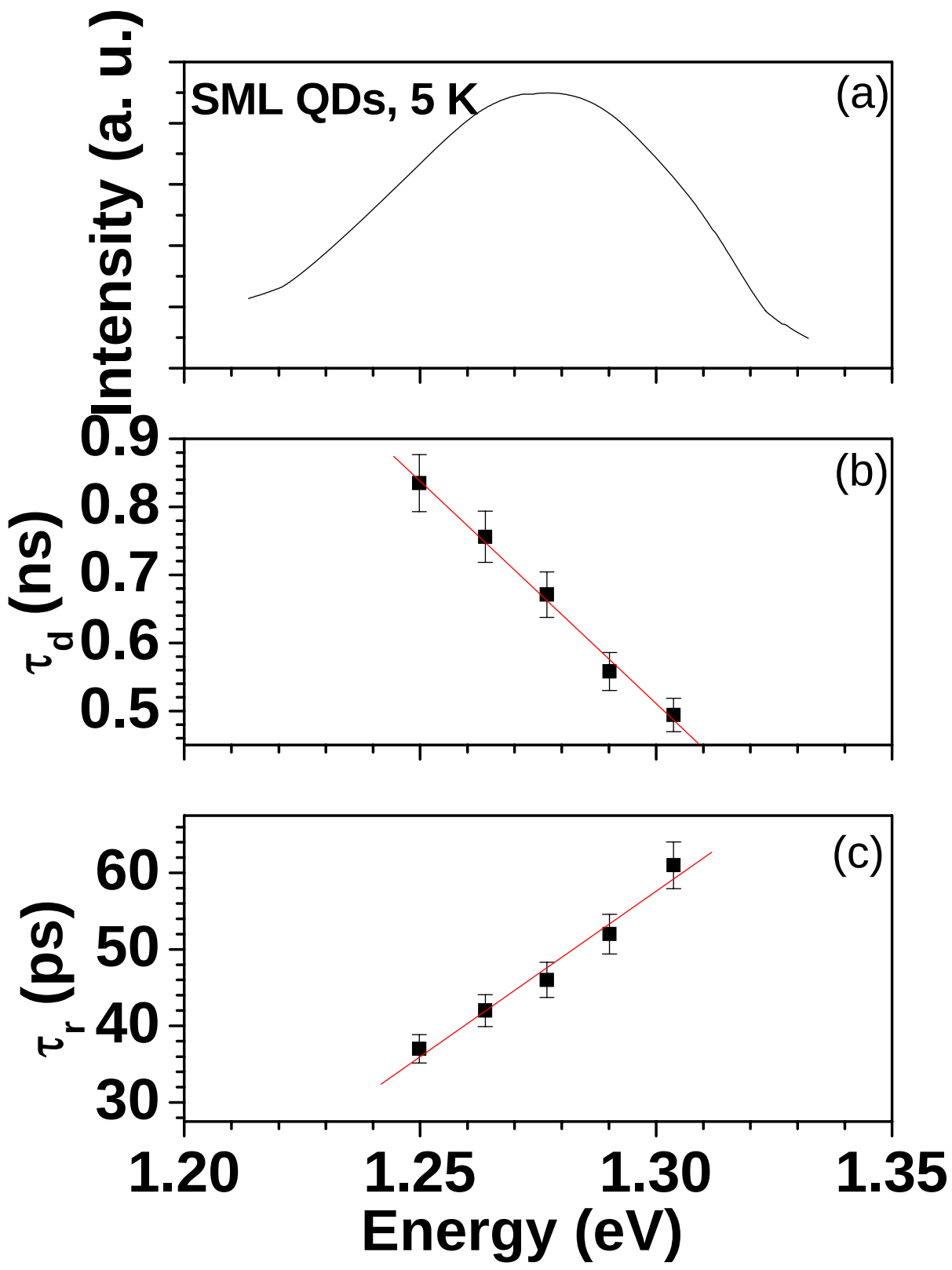

Fig. 4 Zhangcheng Xu et al.

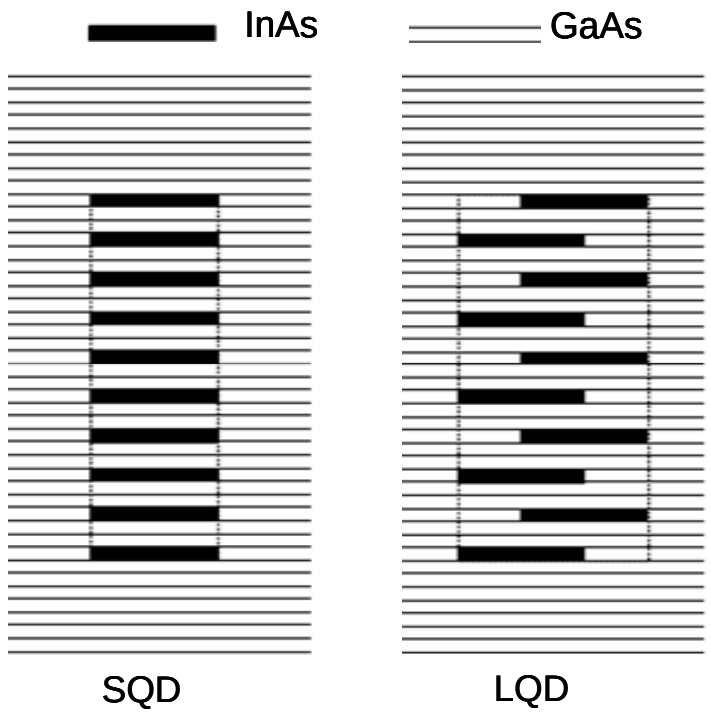

Fig. 5 Zhangcheng Xu et al.